\documentclass[pss]{wiley2sp} % provides pss two-column style
\usepackage{amsmath, amssymb, wasysym}
\begin{document}

% Title of the article
\title{Disordered magnetism 
%phases 
in superconducting KFe$_2$As$_2$ single 
crystals}

% Abbreviated title for the page headers
\titlerunning{Disorded magnetism in superconducting KFe$_2$As$_2$}

% Authors
\author{%
 V.\ Grinenko \textsuperscript{\textsf{\bfseries 1}},
 S.-L.\ Drechsler \textsuperscript{\Ast,\textsf{\bfseries 1}},
  M.\ Abdel-Hafiez \textsuperscript{\textsf{\bfseries 1}},
 S.\ Aswartham \textsuperscript{\textsf{\bfseries 1}},
 A.U.B.\ Wolter \textsuperscript{\textsf{\bfseries 1}},
 S.\ Wurmehl \textsuperscript{\textsf{\bfseries 1}},
 C.\ Hess \textsuperscript{\textsf{\bfseries 1}},
 K.\ Nenkov \textsuperscript{\textsf{\bfseries 1}},
G.\ Fuchs \textsuperscript{\textsf{\bfseries 1}},
D.\ Efremov \textsuperscript{\textsf{\bfseries 1}},
 B.\ Holzapfel \textsuperscript{\textsf{\bfseries 1}},
 J.\ van den Brink \textsuperscript{\textsf{\bfseries 1}},
 B. B\"uchner \textsuperscript{\textsf{\bfseries 1,2}}}

% Abbreviated list of authors for the page headers
\authorrunning{V.\ Grinenko et al.}

%E-mail-address of corresponding author
\mail{e-mail
  \textsf{s.l.drechsler@ifw-dresden.de}, Phone:
  +49-351-4659-384, Fax: +49-351-4659-380}

% author's affiliations/addresses
\institute{
  \textsuperscript{1}\,IFW Dresden, D-01171 Dresden, Germany\\
  \textsuperscript{2}\,Technische Universit\"at Dresden, Institut f\"ur 
Festk\"orperphysik, Germany}

\received{XXXX, revised XXXX, accepted XXXX} % do not change, will be filled in by the publisher
\published{XXXX} % do not change, will be filled in by the publisher

% Please select about four verbal keywords for your manuscript.

\abstract{%
\abstcol{%
 High-quality
 KFe$_{2}$As$_{2}$ (K122) single crystals synthesized by different techniques
have been studied by magnetization and specific heat (SH) measurements. 
The adopted phenomenological
analysis of the normal state properties 
%of dc magnetization and ac susceptibility data 
%are 
%frequently
shows that 
there are two types of samples both
affected by disordered magnetic 
phases: (i) cluster-glass (CG) like or (ii) 
%and 
Griffiths phase (G) like. For (i)  at low
applied magnetic fields the $T$-dependence of the zero field cooled 
(ZFC) linear susceptibility ($\chi_l$) exhibits an anomaly with 
an irreversible 
behavior in ZFC and field cooled (FC) data. This anomaly is related 
to
%with 
the freezing temperature $T_f$ of a CG. For the investigated samples 
the extrapolated $T_f$ to $B=0$ varies between 50~K and 90~K. Below $T_f$ we 
observed a magnetic hysteresis in the field dependence of the isothermal 
magnetization ($M(B)$). The frequency 
shift of the freezing temperature 
$\delta T_f=\Delta T_f/[T_f\Delta(\ln \nu)]\sim 0.05$ has an intermediate 
value, which provides evidence for the formation of  a 
  }{%
  CG-like state in  
the K122 samples of type (i). 
The frequency dependence of their $T_f$ follows a conventional
power-law divergence of critical slowing down 
$\tau=\tau_{0}[T_{f}(\nu)/T_{f}(0)-1]^{-z\nu^{'}}$
with the critical exponent $z\nu^{'}$=10(2) and a relatively 
long characteristic 
time constant $\tau_0$=6.9$\cdotp 10^{-11}$s also supporting a CG behavior. 
The large value of the Sommerfeld coefficient obtained from
SH measurements of these samples was related to magnetic contribution 
from a
CG.
%In contrast, some other K122 
Samples from (ii) did not show a hysteresis 
behavior for 
%in the temperature dependence of  
$\chi_l(T)$ and $M(B)$. 
%On the other hand 
Below 
some crossover temperature $T^{*}\sim$40K a power-law dependence in the 
$\chi_l\propto T^{\lambda_{\mbox{\tiny G}}-1}$,  with a 
non-universal 
$\lambda_{\mbox{\tiny G}}$ was observed, suggesting a quantum 
G-like behavior. 
In this case  $\chi_l$ 
and $M(B)$ can be scaled using the scaling function 
$M_s(T,B)= B^{1-\lambda_{\mbox{\tiny G}}}Y(\mu B/k_{b}T)$ with the 
scaling moment 
$\mu \sim$3.5$\mu_b$. The same non-universal exponent was found also in SH 
measurements, where the magnetic contribution 
$C/T\propto T^{\lambda_{\mbox{\tiny G}}-1}$. 
}}

\maketitle   % please do not remove

\section{Introduction}
The interplay between superconductivity (SC) and magnetism, and 
the role of correlation effects in Fe-pnictides are under debate 
\cite{Johonston2010,Stewart2011,Bang2012}.
It is commonly assumed that in most of the
so-called stoichiometric
parent
 compounds nesting
between electron (el) and
hole (h) Fermi surfaces is responsible for the presence of 
 long-range spin density waves (SDW). To get  SC,
 the SDW should 
be suppressed by chemical doping or external pressure 
\cite{Johonston2010,Stewart2011}. Therefore, it is believed that SC is 
driven at least partially
by AFM spin fluctuations (sf).
In contrast,
in some other of Fe-pnictides such as LiFeAs (111) or 
KFe$_2$As$_2$ (K122) there is no
nesting. Therefore, the role of magnetism in the 
SC pairing 
of these
 compounds is less obvious but it is still believed that remaining  
sf 
as in 
Ba122 \cite{Hirano12} or a new type \cite{Zhang2010} sf can be 
responsible for SC. 

In this paper we demonstrate that the
physical properties 
of clean 
K122 single crystals are strongly affected by an unexpected glassy like 
magnetic 
behavior
such as
spin-glass (SG) and Griffiths (G) type. It is known that a SG phase 
gives a nearly 
linear magnetic contribution to the SH below the freezing temperature $T_f$
\cite{Binder1986,Mydosh1993}. In some cases the SG contribution can be hardly 
distinguished from the usual  
electronic (Sommerfeld) contribution to the SH, since it has the same 
linear $T$-dependence and only a weak maximum slightly above $T_f$. 
Therefore, the Sommerfeld coefficient and the
deduced strength 
of correlation effects can be signifiantly
overestimated, 
if one considers only SH data ignoring thereby the glassy phases. Moreover, the 
interplay of 
superconductivity 
(SC) and {\it unknown} magnetic phases can lead to  confusing conclusions 
concerning SC gaps \cite{Kim2011}. A clear 
understanding of
various coexisting or competing forms of magnetism
should be addressed first.

\section{Experimental results and discussion}

\subsection{Samples}
K122 single crystals have been grown using a
self-flux method (FeAs-flux (S1) and KAs-flux (S2)).
The high quality of the grown single crystals was
assessed by complementary techniques. Several samples were
examined with a Scanning Electron Microscope (SEM Philips XL 30)
equipped with an electron microprobe analyzer for a
semi-quantitative elemental analysis using the energy dispersive
x-ray (EDX) mode (for details see 
\cite{Hafiez2011,Aswartham2012}). Resistivity of all measured samples
showing a metallic behavior at all $T$ with a 
RRR$_5=\rho({\rm300\,K})/\rho ({\rm5\,K})\approx$400-500, where 
$\rho({\rm300\,K})$ and $\rho ({\rm5\,K})$ is resistivity at $T=300$ and 5~K, 
respectively, comparable with the best values reported in the literature. 
The resistivity data 
will be published elsewhere.
Low-$T$ SH and ac susceptibility were 
determined using a PPMS (Quantum Design).
The dc magnetic susceptibility has  been measured in a SQUID 
(Quantum Design) magnetometer. In this paper the data for two 
representative single crystals are shown. 

\subsubsection{Magnetization measurements}
\begin{figure}[t]
\includegraphics[width=18pc,clip]{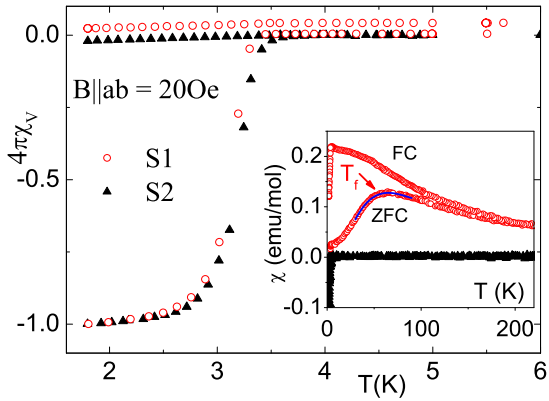}
\caption{(Color online) The temperature dependence of the volume susceptibility 
of S1 and S2 in the 
SC state.
Inset: Their molar susceptibilities in the normal state.} 
\label{Fig:1}
\end{figure}

\paragraph{Samples with cluster glass behavior}

\begin{figure}[t]
\includegraphics[width=18pc,clip]{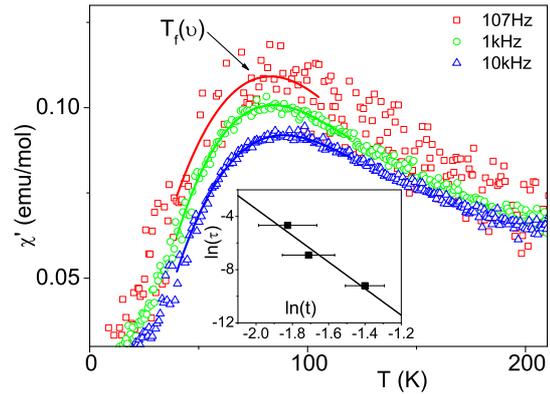}
\caption{(Color online) $T$-dependence of the real part of
the ac susceptibility for sample S1 measured for three 
different frequencies $\nu$  at 5~Oe ac field amplitude.
Inset: the $\nu$-dependence of the 
freezing
temperature plotted as  $ln(\tau)$ vs.\ $\ln (t)$, where $t$ denotes the 
reduced temperature 
$t=T_f(\nu)/T_f(0)-1$.} 
\label{Fig:3}
\end{figure}

Fig.\ \ref{Fig:1} depicts the $T$-dependence of 
the volume susceptibility ($\chi_{v}$) determined from dc magnetization
of our samples measured under both
zero-field-cooled (ZFC) and field-cooled (FC) conditions with the
field $B_{\parallel ab} = 20$\,Oe. Bulk SC of our samples
is confirmed by 'full' diamagnetic signals of the ZFC data
at low $T$. For sample S1, a clear splitting  between ZFC and FC normal 
state linear susceptibility $\chi_l(T)=M(T)/B$ curves is observed below 100~K 
(see the inset of Fig.~\ref{Fig:1}), where $M(T)$ is a magnetization 
in the field $B$. 
The maximum in the ZFC $\chi_l(T)$ is attributed to the
freezing temperature of a spin glass (SG) type phase 
at $T_f \approx 65$~K and $B=20$~Oe \cite{remres1}.
\begin{figure}[t]
\includegraphics[width=18pc,clip]{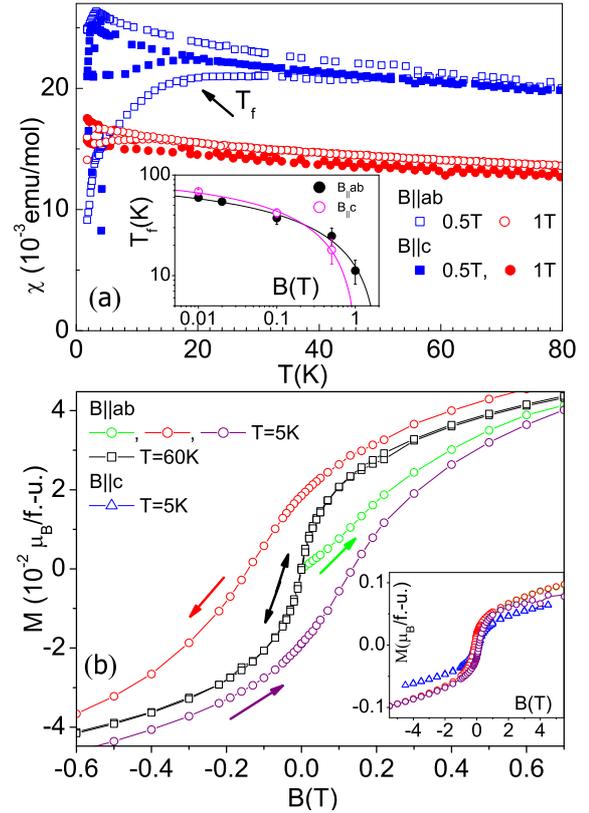}
\caption{(Color online) (a) The molar susceptibility
$\chi_l(T)=M/B$ for S1 crystal measured at different magnetic fields $B$.  
$M$ is the magnetization. Inset: field 
dependence of freezing temperature for field $B_{\parallel ab}$ and 
$B_{\parallel c }$.
(b) Field dependence of magnetization of S1 crystal measured after 
ZFC at $T$=5K and $T$=60K ($B_{\parallel ab}$) . Inset: field dependence of 
magnetization at $B_{\parallel ab}$ and $B_{\parallel c }$ and $T$=5K.} 
\label{Fig:2}
\end{figure}
$T_f$ decreases with increasing field and at 5~T no 
splitting was observed down to 2~K (see Fig.\ \ref{Fig:2}a). 
The field dependence of  $T_f$ is shown in the inset of Fig.\ \ref{Fig:2}a. 
The extrapolated value of $T_f(B=0)\sim 90$~K  
is a relatively high value. This might point to a large 
concentration of the involved
magnetic moments (MM)  
$\gtrsim$10\% in sample S1 \cite{Mydosh1993}. Such a high value of MM is 
expected from entropy estimations, too (see section 2.2). On the other hand, 
structural investigations did not reveal any impurity phase (see section 2.3).
Therefore, we speculate that the high value of $T_f$ might be  caused
 by a low-lying excited 
incommensurate spin density wave state \cite{Overhauser1960}. For a
more detailed consideration of
this scenario see Ref.\ 13.
%\cite{Grinenko2012}. 
In addition, an upshift of the maximum and its lowering 
with increasing frequency $\nu$ of the ac susceptibility, 
generic for a 
SG ordering \cite{Binder1986,Mydosh1993} 
(Fig.\ \ref{Fig:3}), was observed for crystal S1. 
The value of the frequency shift of $T_f$ \cite{remres1}:
\begin{equation}\label{eq1} 
\delta T_f=\frac{\Delta T_f}{[T_f\Delta(log \nu)]}\sim 0.05 .
\end{equation} 
is above the range 0.001-0.02 
expected for canonical SG but well below $\sim$ 0.3 observed 
in the case of superparamagnets \cite{Mydosh1993}. Such an
intermediate value of 
the frequency shift is usually related to a the so-called cluster glass (CG) 
behavior \cite{Marcano2007,Anand2012}. The frequency dependence of the $T_f$ 
shown in inset Fig.~\ref{Fig:3} follows a conventional power-law divergence
of critical slowing down \cite{Mydosh1993,Anand2012}:
\begin{equation}\label{eq2} 
\tau=\tau_{0}[\frac{T_{f}(\nu)}{T_{f}(0)}-1]^{-z\nu^{'}} ,
\end{equation}
where $\tau=1/\nu$ is the relaxation time corresponding to the measured
frequency $\nu$, $\tau_0$ is the characteristic relaxation time of
single spin flip, $T_f(\nu=0, B=5 Oe)\approx$71~K is the spin-glass 
temperature as the  
frequency tends to zero adopted from dc susceptibility measurements 
(inset Fig.~\ref{Fig:2}a), and $z\nu^{'}$ is the dynamic critical exponent. 
It is convenient to rewrite
Eq.~\ref{eq2} in the form:
\begin{equation}\label{eq2a} 
ln(\tau)=ln(\tau_0)-z\nu^{'}ln(t),
\end{equation} 
where $t=T_f(\nu)/T_f(0)-1$. The fit of the
experimental data by a power-law 
divergence  Eq.~\ref{eq2a} is shown in the 
inset Fig.~\ref{Fig:3}. The best fit 
was obtained with $z\nu^{'}$=10(2) and $\tau_0=6.9\cdotp 10^{-11}$s. 
The value of $z\nu^{'}$ is in the range of 4 to 12 observed for typical SG 
\cite{Anand2012}. On the other hand the value of   $\tau_0$ is large as 
compared to $10^{-12} - 10^{-14}$s
observed for structureless SG systems, which is of the order of the spin-flip 
time of atomic magnetic moments ($10^{-13}$s) \cite{Malinowski2011}. 
This suggests a slow spin dynamics in our crystal S1, likely due to the presence of 
interacting clusters rather than individual spins.
  
Another signature of a SG-like behavior for  crystal S1 is a hysteresis 
in the 
magnetization data below $T_f$ 
with a reduced ZFC susceptibility at low fields (Fig.\ \ref{Fig:2}b). 
This behavior is expected in the case of SG or CG systems 
\cite{Mydosh1993,Marcano2007,Coles1978} 
below $T_f$ and also excludes superparamagnetic behavior in our samples 
where no
hysteresis was observed \cite{Bean1959}. 
On the other hand, in the case of canted antiferromagnetic (AF) 
or ferromagnetic (FM) impurity phases hysteresis is expected but with 
a higher susceptibility at low fields because the 
clusters are at first saturated 
along their local easy axis, and only after that various clusters
become fully aligned along the applied field \cite{Malinowski2011}. 
Therefore, our $M(B)$ data exclude
(large) clusters of impurity phases such as FeAs, Fe$_2$As or other iron 
compounds. The same conclusion can be drawn
from our SH measurements (see below). 
We also observed the displacement of the 
ZFC magnetization compared to magnetization after complete hysteresis loop at 
magnetic field applied parallel to $ab$ plane
(inset Fig.~\ref{Fig:2}b). For the field along the
$c$-axis no 
displacement was observed. This indicates that the glassy behavior 
is dominated by a magnetic interaction between moments lying in the 
$ab$-plane. 
        
\paragraph{Samples with Griffiths-phase behavior}

\begin{figure}[t]
\includegraphics[width=18pc,clip]{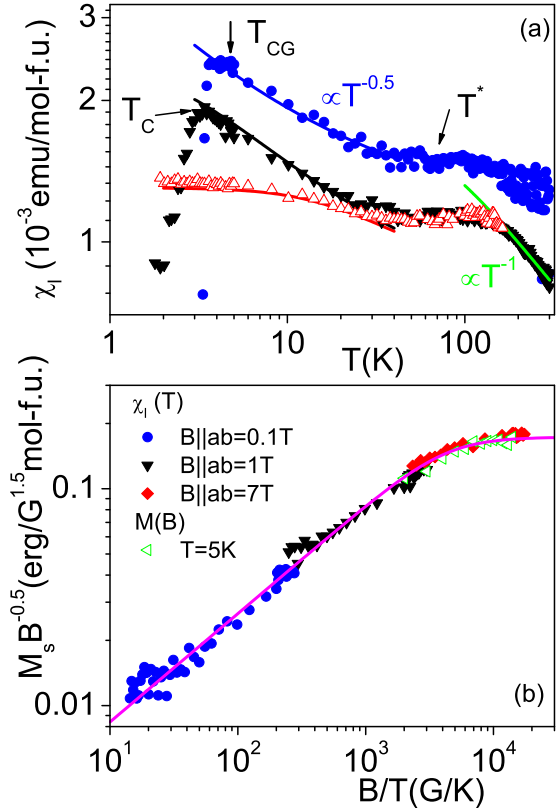}
\caption{(Color online) (a) The molar susceptibility
$\chi_l(T)=M/B$ for crystal S2 measured in
 different magnetic fields $B$, where 
$M$ is the magnetization. Fitting curves using Eq.~\ref{eq3a}. 
(b) The scaled magnetization $M_{s}B^{-0.5}=(M(T,B)-\chi_{l0}B)B^{-0.5}$ vs.\ 
$B/T$ for crystal S2.(see Eq.~\ref{eq3a})} 
\label{Fig:4}
\end{figure}

In contrast, the $T$-dependence of the linear susceptibility  $\chi_l(T)$ of 
some crystals S2
does not show a 
difference 
between ZFC and FC curves above $T_c$. The $\chi_l(T)$ data of one of the S2 
crystals is shown in Fig.~\ref{Fig:4}a. 
At high $T>$200~K, $\chi_l(T)$ follows a Curie-Weiss behavior with an AFM 
$\Theta_c$=-117~K \cite{remres2}. 
At $T\lesssim$120~K $\chi_l(T)$ shows a plateau-like feature with a field 
independent susceptibility above $B=1$~T. In a view of the 
observed CG in  
sample S1, we relate this flattening to a tendency to form magnetic 
clusters in the
crystals S2, too. However, with lowering $T$ after a weak reduction of 
$\chi_l(T)$ instead to form a CG phase in crystal S2, 
below $T^*$ an exponential increase of the susceptibility is observed:
\begin{equation}\label{eq3}
\chi_{l}(T)= \chi_{l0}+\frac{Cu_{\mbox{\tiny G}}}{T^{1-\lambda_{\mbox{\tiny G}}}},
\end{equation}         
where $\chi_{l0}$ is a $T$-independent susceptibility, 
and $Cu_{\mbox{\tiny G}}$ is a back ground constant. 
A power-law with the exponent 
$\lambda_{\mbox{\tiny G}}$ was found up to the 
highest measured field 7~T (see Fig.\ \ref{Fig:4}a).
This exponential behavior is quite similar to the 
reported one for the weak itinerant ferromagnetic alloy Ni$_{1-x}$V$_x$ 
\cite{Ubaid-Kassis2010}, where 
the formation of a Griffiths (G) phase with a 
non-universal exponent was 
observed near a FM quantum critical point. Following the analysis 
proposed there, the 
field and $T$-dependence of the magnetization can be scaled on a single
 curve 
Fig.~\ref{Fig:4}b: 
\begin{equation}\label{eq3a}
M_s(T,B)= B^{1-\lambda_{\mbox{\tiny G}}}Y(\frac{\mu B}{k_{b}T}),
\end{equation}
where $\mu$ is the scaling moment and 
$Y=A^{'}/(1+z^{-2})^{\lambda_{\mbox{\tiny G}}/2}$ a
scaling function with 
$A^{'}=A/\mu^{\lambda_{\mbox{\tiny G}}}$ as a constant. 
To scale the data using Eq.~\ref{eq3a} we have subtracted 
the $T$-independent susceptibility 
$\chi_{l0}$ from  $\chi_{l}(T)$ and $\chi_{l0}B$ from $M(H)$, correspondingly.
For sample S2 (Fig.~\ref{Fig:4})  a scaling was observed for 
$\lambda_{\mbox{\tiny G}} \approx$0.5(1) 
with a scaling moment $\mu \sim$3.5$\mu_b$. According to Ref.\ 20
%\cite{Ubaid-Kassis2010} 
the obtained moment can be related 
to a typical cluster size in crystal S2. 
The SH data are also in agreement with the G-like 
scenario (see below). Therefore, we ascribe the anomalous 
power-law of $\chi(T)$ at low $T<T^*\approx 50$~K to the 
formation of a 
quantum G-phase.

\subsection{Specific heat measurements}

\paragraph{Specific heat in the normal state}

\begin{figure}[t]
\includegraphics[width=21pc,clip]{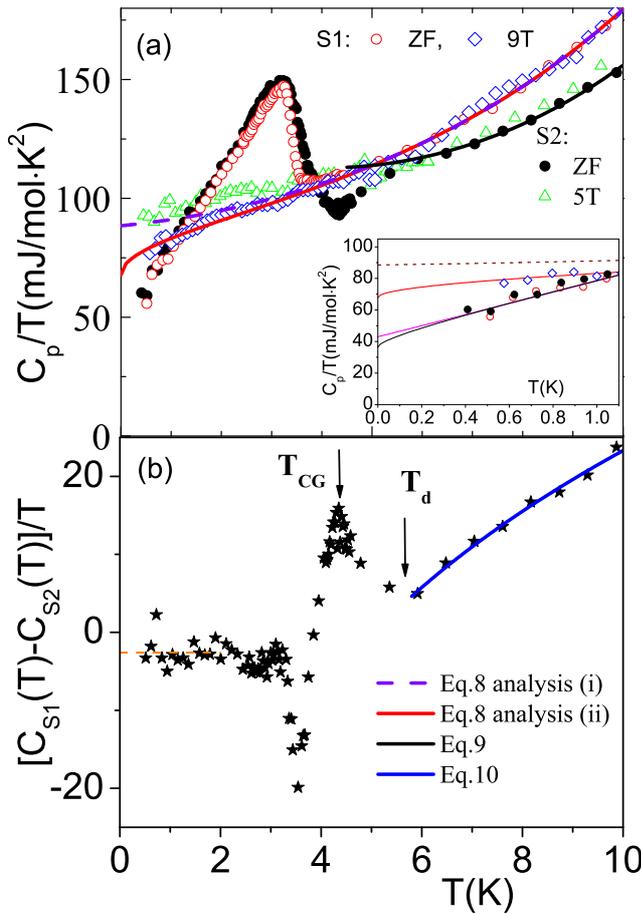}
\caption{(Color online) (a) The specific heat of our two K122 
samples. Inset: low temperature part of the SH at zero field.
(b) The plot $[C_p^{S1}(T)-C_p^{S2}(T)]/T$.}
\label{Fig:5}
\end{figure}

We have found that the glassy magnetic subsystems (observed in the magnetization 
measurements) do 
also contribute to the 
SH shown in Fig.~\ref{Fig:5}a. In case of SG or CG 
phases the magnetic contribution
$C_{CG}$ to the SH varies almost linearly at $T<T_f$ 
like the usual electronic contribution in case of a FL
\cite{Mydosh1993,Binder1986,Dawes1979,Martin1980}. Empirically, 
this behavior can be approximated by 
\begin{equation}\label{eq4a}
C_{\rm CG}\approx \gamma_{\rm CG}T+\varepsilon_{\rm CG2}T^2,
\end{equation} 
or
\begin{equation}\label{eq4b}
C_{\rm CG}\approx 
%\gamma_{\rm CG}T+
\varepsilon_{\rm CG1.5}T^{1.5},
\end{equation}
where $\gamma_{\rm CG}$, $\varepsilon_{\rm CG2}$ and $\varepsilon_{\rm CG1.5}$
are CG constants. The  $\varepsilon_{\rm CG1.5}$ contribution can be interpreted 
as originating from short-range 3D ferromagnetic (FM) spin waves which can 
exist in a FM clusters \cite{Thomson1981}, also a linear contribution to SH can 
be expected for 2D FM spin waves. 
Then, the normal state SH of sample S1 reads 
\begin{equation}\label{eq4d}
C_p^{S1}(T)=\gamma_{\rm el}T+C_{\rm m}^{CG}+\beta_3 T^3+\beta_5 T^5 \ ,
\end{equation}
where $C_{\rm m}^{CG}$ is given by Eq.~\ref{eq4a} and Eq.~\ref{eq4b}, 
$\gamma_{\rm el}$ is an intrinsic electronic contribution and 
$\beta_3$, $\beta_5$ are a lattice contribution.
In case of a G-phase (sample S2), 
$C(T)_G/T \propto \chi(T)$ is expected \cite{CastroNeto1998,Stewart2001}. 
Hence, for the SH we have:
 \begin{equation}\label{eq5}
C_p(T)^{S2}=\gamma_{\rm el}T+\gamma_GT^{\lambda_G}+\beta_3 T^3+\beta_5 T^5 \ ,
\end{equation}
where $\lambda_{\mbox{\tiny G}} \approx 0.5(1)$ according to our magnetization 
data. 
To reduce the number of fitting parameters in Eqs.~\ref{eq4d} and \ref{eq5}, 
we analyzed the difference:
\begin{equation}\label{eq6}
[C_p^{S1}(T)-C_p^{S2}(T)]/T=
%\Delta\gamma_{\rm el}+
C_{\rm CG}^{S1}-\gamma_GT^{\lambda_G-1}\ .
\end{equation}
This allow us to exclude the lattice contributions $\beta_3$, $\beta_5$,
as well as the linear electronic term $\gamma_{\rm el}$
which are all
supposed to be nearly the same for both crystals, respectively. The fit of the 
experimental data by Eq.~\ref{eq6} is shown in Fig.~\ref{Fig:5}b. 
i) In the case of  Eq.~\ref{eq4a} it gives:  
%$\Delta\gamma_{\rm el}+
$\gamma_{\rm CG}^{S1}=36$~mJ/mol$\cdotp$K$^2$, 
$\varepsilon_{\rm CG2}=2.0$~mJ/mol$\cdotp$K$^3$ and 
$\gamma_{\rm G}=104$~mJ/mol$\cdotp$K$^{1.5}$, respectively. 
Then, using in Eq.~\ref{eq5} obtained magnetic contribution
we have estimated the {\it intrinsic}  
$\gamma_{\rm el}^i\approx 52$~mJ/mol$\cdotp$K$^2$ for sample S2
with the lattice terms
$\beta_3^i=0.59$~mJ/mol$\cdotp$K$^4$ and 
$\beta_5^i=1.20\cdotp 10^{-3}$~mJ/mol$\cdotp$K$^6$, 
respectively. The obtained $\beta_3^i$ corresponds to a
Debye temperature 
$\Theta_D^i\approx 254$~K.
ii) In the case of  validity of Eq.~\ref{eq4b} it gives:  
%$\Delta\gamma_{\rm el}+\gamma_{\rm CG}^{S1}\approx 0$~mJ/mol$\cdotp$K$^2$, 
$\varepsilon_{\rm CG1.5}=14.9$~mJ/mol$\cdotp$K$^3$ and 
$\gamma_{\rm G}=75.3$~mJ/mol$\cdotp$K$^{1.5}$, respectively. 
Then the {\it intrinsic} 
$\gamma_{\rm el}^{ii}\approx 68$~mJ/mol$\cdotp$K$^2$ is the same for S1 and S2
 crystals with slightly different lattice terms as compared to those
  obtained in the  
analysis (ii): $\beta_3^{ii}=0.46$~mJ/mol$\cdotp$K$^4$ and 
$\beta_5^{ii}=1.86\cdotp 10^{-3}$~mJ/mol$\cdotp$K$^6$,
respectively. This value of $\beta_3^{ii}$ corresponds to a Debye temperature 
$\Theta_D^{ii}\approx$276K. Both analysis give reasonable values of the lattice 
contribution (for example, in the case of Ba$_{0.68}$K$_{0.32}$Fe$_{2}$As$_{2}$ 
a Debye temperature of 277K was estimated \cite{Popovich2010}) and essentially 
reduced $\gamma_{\rm el}\approx 52-68$~mJ/mol$\cdotp$K$^2$ as compared to 
nominal values $\gamma_{\rm n}\approx 100$~mJ/mol$\cdotp$K$^2$ 
\cite{Fukazawa2011,Kim2011} obtained with out accounting for 
 magnetic contributions. The SH data at
$B=9$~T shown in Fig.~\ref{Fig:5}a can be considered as a support of 
analysis (ii), since this analysis provides essentially a better agreement 
between the  
experimental data and the fitting curves at low-$T$. However, we cannot 
exclude that large field enhances FM order in S1 crystals and actually change 
the entropy of SG at low temperatures.     
      
\begin{figure}[b]
\includegraphics[width=21pc,clip]{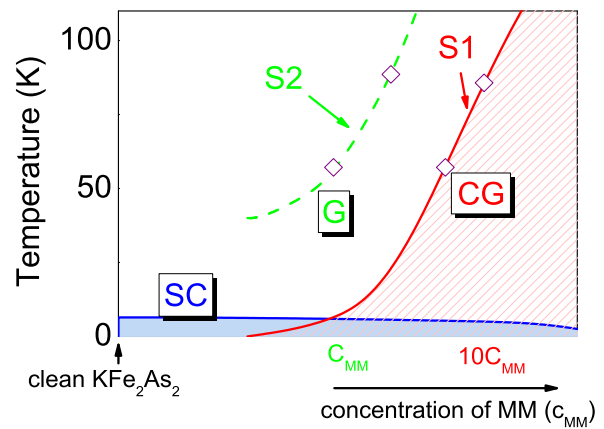}
\caption{(Color online) Schematic phase diagram of  
extrinsic magnetic moments (MM) driven 
quantum phase transitions. 
Notation of phases: SC - superconductivity, G - Griffiths, CG - cluster glass, 
S1, S2 - samples from this work. 
%H - from Ref.\ 29
%~\cite{Hashimoto2010} 
%and 
%D - from Ref.\ 30.}
%~\cite{Dong2010}
}
\label{Fig:6}
\end{figure}

Below $T_d$ $\sim$6~K the data for
S2 deviate from the fit-curve (Fig.~\ref{Fig:5}).
At $T_{\rm CG} \sim$4~K, slightly above $T_c$, (see 
Fig.\ \ref{Fig:5}a) another magnetic anomaly is well visible in the
SH data. Additionally, slightly above $T_{\rm CG}$ we observed a plateau-like 
feature in $\chi(T)$ at relatively low fields (see Fig.~\ref{Fig:4}a). 
We ascribe $T_d$ to the freezing of large cluster
dynamics in accord with the behavior expected for a 
quantum G-phase 
(see Ref.~\cite{Vojta2010}) followed by the final
formation of a CG phase due to the RKKY interaction between 
the 
clusters at $T_{\rm CG}$
in crystal S2, too (for an illustration see also Fig.~\ref{Fig:6}).

\paragraph{Specific heat in the superconducting state}

Measurements in SC state have shown that there is a large 
residual Sommerfeld coefficient observed for all investigated samples 
(see inset Fig.~\ref{Fig:5}a). The fit below 1K gives a residual 
contribution for crystal S1
$\gamma_{\rm res1}^{S1}\approx$43~mJ/mol$\cdotp$K$^2$ and about 
$\gamma_{\rm res1}^{S2}\approx$46~mJ/mol$\cdotp$K$^2$ for S2 
\cite{remminorphases}. The  $\gamma_{\rm res1}^{S1}$  
is close to 
$\gamma_{\rm CG}^{S1}=36$~mJ/mol$\cdotp$K$^2$ 
estimated for the normal state using 
analysis (i). The closeness of these values would indicate that 
$\gamma_{\rm SG}^{S1}$ is weakly 
effected by the SC transition and also excludes essentially a
 non-superconducting 
volume fraction for  
our the samples. The latter is also supported by the large absolute value of 
the SH jump at $T_c$ compared to the reported in the literature values 
\cite{Kim2011,Fukazawa2011}. 
In the case of \cite{Kim2011} it was observed that $C_p/T$ of the 
investigated crystals tends to zero at $T=0$ after AFM type magnetic 
transition at $T\sim$0.7K. This demonstrates that almost all itinerant 
quasi-particals are gapped at $T=0$. Therefore, we conclude that the large 
residual $\gamma_{\rm res}$ in the SC state of our samples is mainly due 
to the magnetic 
contribution from a CG. On the other hand, using 
$\varepsilon_{\rm CG1.5}=14.9$~mJ/mol$\cdotp$K$^3$ from analysis (ii),
we get $\gamma_{\rm res2}^{S1}\approx$36~mJ/mol$\cdotp$K$^2$. This value is  
nearly a half of $\gamma_{\rm el}^{ii}\approx 68$~mJ/mol$\cdotp$K$^2$. 
In contrast to the conclusion obtained from analysis (i), it 
would mean that the CG phase in SC state is different from the CG in 
the normal state,
since we exclude a large non-SC part of our samples. This can be possible, since 
itinerant electrons responsible for the RKKY interaction  
are affected by the SC transition. Thus, on this stage we cannot decide which 
analysis (i) or (ii) is more sophisticated. Therefore, we estimate the
{\it intrinsic} $\gamma_{\rm el}^{ii}\approx 52-68$~mJ/mol$\cdotp$K$^2$.
A more detailed report of the superconducting properties
including microscopic considerations will be given elsewhere.  

\subsection{Possible disorder induced quantum phase transitions}

Up to now the structural investigation of the 
cleaved surface of the samples such as EDX, XRD 
and SEM did not reveal any secondary phases \cite{Hafiez2011,Aswartham2012}. 
Therefore, we enforced to adopt a 'point'-defect model such as vacancies or 
interstitials of Fe atoms.  
To compare the amount of magnetic clusters contributing to a glassy
phases of our samples, we calculated the magnetic entropy 
$S_{m}=\varint(C_{m}/T)dT$ using the obtained above magnetic contributions. 
$S_m^{S2}$ for crystal S2 related to CG and G phases between 0 and 
$T^*\approx 50$~K 
(where the quantum 
Griffiths behavior appears in the 
magnetization data Fig.~\ref{Fig:4}a) 
is $\sim$0.074RJ/mol-Fe$\cdotp$K$^2$. The estimate for crystal S2
related to the CG phases below $T_{f}\approx 87$~K gives an essentially 
higher value than $S_m^{S1}$. In the case of validity of
analysis (i) $\sim$0.64RJ/mol-Fe$\cdotp$K$^2$ and for the case of 
analysis (ii)
$\sim$0.48RJ/mol-Fe$\cdotp$K$^2$ for the magnetic contribution 
have been
obtained, respectively. 
Hence, we conclude that in the normal state  crystal S1, it can contain 
up to 10 times more magnetic clusters than S2 does. 
Summarizing our experimental 
observations of disordered magnetic phases in K122, we can propose  
a quantum phase transition of spin glass type with strong quantum G-phase 
effects (see Fig.~\ref{Fig:6}) driven by some tuning 
parameter $p$ which is responsible for the formation of magnetic moments 
(MM) in K122. The physical nature of $p$ should be unraveled in future 
investigations such as spin and nuclear magnetic resonance and/or M\"ossbauer 
spectroscopy. These techniques can be helpful to estimate the amount and the 
distribution of MM in K122 single crystals. 

\section{Conclusions}
 To summarize,
analyzing magnetization and specific heat  
data, we found out that even in high-quality
KFe$_2$As$_2$ single crystals
glassy magnetic behavior like in spin- , cluster-glasses 
and Griffiths phases
may occur near superconductivity and 
%even 
coexist with it. 
The magnetic contribution is responsible for a large value of the nominal 
Sommerfeld coefficient 
$\gamma_{n}\sim$100~mJ/mol$\cdotp$K$^2$ of this compound. 
The analysis of the SH data has 
shown that magnetic contribution amounts up to 50$\%$ of $\gamma_{n}$. 
In this way, the intrinsic value of the Sommerfeld coefficient 
$\gamma_{\rm el}\approx 52-68$~mJ/mol$\cdotp$K$^2$ was estimated. We observed 
that various samples exhibit different disordered magnetic contributions 
depending 
on the amount and distribution of magnetic moments (MM). 
This suggests an
extrinsic origin of MM which can be caused by point defects such as 
vacancies or Fe 
interstitials. Therefore, we proposed a scenario of disorder induced spin 
glass type quantum phase transitions accomplished by strong quantum Griffiths 
effects. Further investigations are required to elucidate the physical 
origin 
and the distribution of such MM.  
 
\begin{acknowledgement}
We thank
J.A.\ Mydosh,
U.\ R\"o{\ss}ler, and
D.\ Evtushinsky
for 
discussions. 
Our work
was supported by the DFG (SPP 1458
and the Grants No. GR3330/2, WU 595/3-1 (S.W.)) 
as well as the IFW-Pakt f\"ur Forschung.
\end{acknowledgement}


\begin{thebibliography}{[1]}
\bibitem{Johonston2010}%
D.\,C. Johnston Advances in Physics \textbf{59}, 803 (2010).
\bibitem{Stewart2011}%
G.\,R. Stewart,  Rev.\ Mod.\ Phys.\ \textbf{83}, 1589 (2011).
\bibitem{Bang2012}%
Y. Bang Supercond.\ Sci.\ Technol. \textbf{25}, 084002 (2012).
\bibitem{Hirano12}%
M.\ Hirano, Y.\ Yamada, T.\ Saito {\it et al.}
%, R.\ Nagashima, T.\ Konishi, T.\ Toriyama, 
%Y.\ Ohta, H.\ Fukazawa, Y.\ Kohori, Y.\ Furukawa, K.\ Kihou, C.-H.\ Lee, 
%A.\ Iyo, and H.\ Eisaki, 
J.\ Phys.\ Soc.\ Jpn.\ \textbf{81}, 054704 (2012).
\bibitem{Zhang2010}%
S.\,W. Zhang, L.\ Ma, Y.\,D.\ Hou {\it et al.}
%, J.\ Zhang, T.-L.\ Xia, G.\,F.\ Chen, 
%J.\,P.\ Hu, G.\, M.\ Luke and W.\ Yu, 
Phys.\ Rev.\ B \textbf{81}, 012503 (2010).
\bibitem{Binder1986}%
K.\ Binder and A.\,P. Young, Rev.\ Mod.\ Phys.\ \textbf{58}, 801 (1986).
\bibitem{Mydosh1993}%
J.A.\ Mydosh, Spin glasses: an experimental introduction 
(Taylor and Francis, London - Washington, DC, 1993).
\bibitem{Kim2011}%
J.\,S.\ Kim, E.\,G.\ Kim, G.\,R.\ Stewart, X.\,H. Chen, and X.\,F. Wang,  
Phys.\ Rev.\ B  \textbf{83}, 172502 (2011).
\bibitem{Hafiez2011}%
M.\ Abdel-Hafiez , S.\ Aswartham, S.\ Wurmehl, V.\ Grinenko, 
S.-L.\ Drechsler, 
A.U.B.\ Wolter, and B.\ B\"{u}chner Phys.\ Rev.\ B {\bf 85}, 134533 (2012).
\bibitem{Aswartham2012}%
S.\ Aswartham, S.\ Wurmehl, and B.\ B\"{u}chner  unpublished.
\bibitem{remres1}%
To define the $T_f$ by the same criteria, the $T$-dependence of 
the
dc and the ac susceptibilities were fitted by a cubic polynom
as shown in Figs.~\ref{Fig:1}-\ref{Fig:3}. Next $T_f$ was 
attributed to the maximum calculated from that cubic 
polynom.
\bibitem{Overhauser1960}A.W.\ Overhauser,
%Phys.\ Rev.\ Lett.\ {\bf 3}, 414 (1959);
J.\ Phys.\ Chem.\ Solids {\bf 13} 71 (1960).
\bibitem{Grinenko2012} V.\ Grinenko {\it et al.}, e-print arXiv:1203.1585.
\bibitem{Marcano2007}%
N.\ Marcano, J.\,C. Gomez Sal, J.\,I. Espeso, L.\ Fernandez Barquin, and 
C.\ Paulsen, Phys.\ Rev.\ B \ \textbf{76}, 224419 (2007).
\bibitem{Anand2012}%
V.\,K. Anand, D.\,T. Adroja, and A.\,D. Hillier, Phys.\ Rev.\ B \ \textbf{85}, 014418 (2012).
\bibitem{Malinowski2011}%
A.\ Malinowski, V.L.\ Bezusyy, R.\ Minikayev, P.\ Dziawa, Y.\ Syryanyy, 
and M.\ Sawicki, Phys.\ Rev.\ B \ \textbf{84}, 024409 (2011).
\bibitem{Coles1978}%
B.\,R. Coles, B.\,V. Sarkissian and R.\,H.\ Taylor Phil.\ Mag. B \textbf{37},
 489 (1978).
\bibitem{Bean1959} C.\,P. Bean and J.\,D. Livingston , 
J.\ Appl.\ Phys.\ \textbf{30}, S120 (1959).
\bibitem{remres2}%
We also observed a Curie-Wiess behavior at high-$T$ for samples with a 
CG-phase having lower values of $T_f$ than shown in Fig.~\ref{Fig:1} and Fig.~\ref{Fig:2}.
\bibitem{Ubaid-Kassis2010}%
S. Ubaid-Kassis, T. Vojta, and A. Schroeder, Phys.\ Rev.\ Lett.\ \textbf{104}, 066402 (2010).
\bibitem{Dawes1979}%
 D.\,G. Dawes and B.R.\ Coles, J.\ Phys.\ F: Metal Phys.\ \textbf{9}, L215 (1979).
\bibitem{Martin1980}%
D.\,L. Martin, Phys.\ Rev.\ B  \textbf{21}, 1906 (1980).
\bibitem{Thomson1981}%
J.\,O. Thomson and J.\,R Thompson J.\ Phys.\ F: Metal Phys.\ \textbf{11}, 247 (1981).
\bibitem{CastroNeto1998}%
A.\,H. Castro Neto, G. Castilla and B.\,A. Jones 
Phys.\ Rev.\ Lett.\ \textbf{81}, 3531 (1998).
\bibitem{Stewart2001}%
G.\,R. Stewart,  Rev.\ Mod.\ Phys.\ \textbf{73}, 797 (2001).
\bibitem{Popovich2010}%
P. Popovich, A.\,V. Boris, O.\,V. Dolgov {\it et al.}
%, A.\,A. Golubov, D.\,L. Sun, 
%C.\,T.\ Lin, R.\,K.\ Kremer, and B.\ Keimer, 
Phys.\ Rev.\ Lett.\  \textbf{105}, 027003 (2010).
\bibitem{Fukazawa2011}%
H.\ Fukazawa, T.\ Saito, Y.\ Yamada {\it et al.}
%, K.\ Kondo, M.\ Hirano, Y.\ Kohori, 
%K.\ Kuga, A.\ Sakai, Y.\ Matsumoto, S.\ Nakatsuji, K.\ Kihou, A.\ Iyo, 
%C.\,H. Lee, and H.\ Eisaki, 
J.\ Phys.\ Soc.\ Jpn.\ \textbf{80}, SA118 (2011).
\bibitem{Vojta2010}%
T.\ Vojta, J.\ Low Temp.\ Phys., \textbf{161}, 299 (2010).
%\bibitem{Hashimoto2010}%
%K.\ Hashimoto, A.\ Serafin, S.\ Tonegawa
%, R.\ Katsumata, R.\ Okazaki, 
%T.\ Saito,H.\ Fukazawa, Y.\ Kohori, K.\ Kihou, C.H.\ Lee, A.\ Iyo, H.\ Eisaki, 
%H.\ Ikeda Y.\ Matsuda, A.\ Carrington, and T.\ Shibauchi, 
%Phys.\ Rev.\ B \textbf{82}, 014526 (2010).
%\bibitem{Dong2010}J.\,K. Dong, S.\,Y. Zhou
%, T.\,Y. Guan, H. Zhang, Y.\,F. Dai, 
%X.\ Qiu, X.\,F. Wang, Y.\ He, X.\,H.\ Chen, and S.\,Y.\ Li, 
%{\it et al.}
%Phys.\ Rev.\ 
%Lett.\ \textbf{104}, 087005 (2010).
\bibitem{remminorphases}%
Note that a large value of $\gamma_{\rm CG}$ is expected 
in case of SG or CG \cite{Mydosh1993,Dawes1979}. In contrast, the magnetic 
impurity phases (large clusters) would give only a small contribution at low T.
For example, Fe$_2$As has $\gamma$=21.8~mJ/mol$\cdotp$K$^2$ \cite{Zocco2012} 
and FeAs has $\gamma\approx 6.65$~mJ/mol$\cdotp$K$^2$ \cite{Gonzalez2012}, only. 
Therefore, even a very high contamination by these impurities would provide 
only several mJ at low T.
\bibitem{Zocco2012}%
 D.A.\ Zocco, D.Y.\ T\"ut\"un, J.J.\ Hamlin {\it et al.}
 %, J.R.\ Jeffries, S.T.\ Weir, 
 %Y.K.\ Vohra, and M.B.\ Maple, 
 Supercond.\ Sci.\ Technol. \textbf{25}, 
 084018 (2012).
\bibitem{Gonzalez2012}%
D.\ Gonzalez-Alvarez, F.\ Gronvold, B.\ Falk, E.F.\ Westrum, R.\ Blachnik, 
and G.\ Kudermann, J.\ Chem.\ Thermodynamics \textbf{21}, 363 (1989).
\end{thebibliography}
\end{document}